# LINA – A social augmented reality game around mental health, supporting real-world connection and sense of belonging for early adolescents

Gloria Mittmann

D.O.T. Research Group for Mental Health of Children and Adolescents, Ludwig Boltzmann Society at Karl Landsteiner University of Health Sciences, Krems, Austria

Department of Developmental and Educational Psychology, University of Vienna, Vienna, Austria, gloria.mittmann@kl.ac.at

Adam Barnard

Self-employed creative practitioner

D.O.T. Research Group for Mental Health of Children and Adolescents, Ludwig Boltzmann Society at Karl Landsteiner University of Health Sciences, Krems, Austria, adamacbarnard@gmail.com

Ina Krammer

D.O.T. Research Group for Mental Health of Children and Adolescents, Ludwig Boltzmann Society at Karl Landsteiner University of Health Sciences, Krems, Austria, ina.krammer@dot.lbg.ac.at

Department of Health and Clinical Psychology, University of Vienna, Vienna, Austria

Diogo Martins

Group of AI for People and Society, Instituto de Engenharia de Sistemas e Computadores: Investigação e Desenvolvimento, Lisbon, Portugal

D.O.T. Research Group for Mental Health of Children and Adolescents, Ludwig Boltzmann Society at Karl Landsteiner University of Health Sciences, Krems, Austria, dg.martins1@gmail.com

João Dias

Faculdade de Ciências e Tecnologia, Universidade do Algarve and CCMAR and INESC-ID, jmdias@ualg.pt

**Abstract**

Early adolescence is a time of major social change; a strong sense of belonging and peer connectedness is an essential protective factor in mental health during that period. In this paper we introduce *LINA*, an augmented reality (AR) smartphone-based serious game played in school by an entire class (age 10+) together with their teacher, which aims to facilitate and improve peer interaction, sense of belonging and class climate, while creating a safe space to reflect on mental health and external stressors related to family circumstance. *LINA* was developed through an interdisciplinary collaboration involving a playwright, software developers, psychologists, and artists, via an iterative co-development process with young people. A prototype has been evaluated quantitatively for usability and qualitatively for efficacy in a study with 91 early adolescents ($age_{mean}$=11.41). Results from the Game User Experience Satisfaction Scale (GUESS-18) and data from qualitative focus groups showed high acceptability and preliminary efficacy of the game. Using AR, a shared immersive narrative and collaborative gameplay in a shared physical space offers an opportunity to harness adolescent affinity for digital technology towards improving real-world social connection and sense of belonging.

# 1 INTRODUCTION

Positive peer relationships and supportive friendships become increasingly important from early adolescence [1]. A sense of belonging – defined as "a pervasive drive to form and maintain at least a minimum quantity of lasting, positive, and significant interpersonal relationships" [2, p. 497] – plays a pivotal role in mental health. Belonging – and school belonging in particular – is an important protective factor during the defining years of school, particularly in early adolescence [3]. There are for example consistent positive relations between a sense of belonging and academic performance, self-concept, behavioural engagement, and a general sense of wellbeing [4-7]. In contrast, victimisation and peer-related stress (e.g., physical aggression, humiliation, exclusion) hinder sense of belonging [8, 9], which often creates peer rejection, resulting in long-term mental health problems [10]. Positive interactions both in real- life and online environments can reduce barriers between peers and improve sense of belonging [9]. Likewise, stigma can be reduced by spending time with 'outsiders' in a safe environment, e.g. undertaking structured, shared and collaborative experiences [11, 12].

In most countries, early adolescence is also the time of transition from primary to secondary school (e.g. at age 10 in Austria, age 11 in the UK), which involves major challenges such as a new and larger school environment, more responsibility and academic pressure and often the necessity to make new friends (when old friends have gone to a different school and because of a substantial influx of new people), which creates a new social dynamic. This transition can be difficult for some early adolescents and lead to a decrease in social and emotional health [13]. Positive peer relations and stable friendships are an important protective factor for successful school transition, while poor peer relations and low-quality friendships serve as a risk factor[14], making early adolescence an especially relevant but also fragile time for sense of belonging, and a crucial window for interventions.

As most adolescent social connections are formed in the school setting [15], classes are an ideal target for interventions promoting sense of belonging and positive peer relations. Traditional social skill trainings (SST) and social and emotional learning (SEL) have been shown to effectively improve adolescents' social-emotional skills and wellbeing [16]. However, adolescent lives are increasingly intertwined with the digital world; friendships are formed and maintained online from early adolescence [17]. Furthermore, most adolescents regularly play and enjoy digital games [81% aged 10-24 report playing games; 18]. Digital technology thus offers opportunities for new ways to implement interventions around SEL and belonging, redeploying adolescent affinity for technology towards supporting real-world peer relations and group dynamics.

Advances in information technology have led to many new approaches to engaging people with educational content in settings that are novel and attractive to young people, and potentially offer pedagogical advantages. Combining the "fun" of games with the "serious" content of education [19] and using well-established motivational components of gamification (such as clear goals and guided paths, or direct feedback and rewards) [20] makes gamified interventions and serious games especially engaging for users and an attractive prospect for both educationalists and psychologists. A specific direction in serious gaming is the use of augmented reality (AR). AR is a technology where virtual components are overlaid onto the real world, so that the player engages with the illusion that this virtual component exists in the physical world. Made famous by the popular commercial game *Pokémon Go* by Niantic in 2016, it has now found its way into educational settings [for literature reviews see 21, 22, 23]. Although rigorous evaluations remain rare, recent literature reviews suggest involvement, engagement, motivational, and (tentatively) pedagogical effects of using AR games for education [21, 24]. There remain many opportunities to improve the impact of serious games. For example, the majority of serious games are not designed by a multidisciplinary team [25].

Our aim was to design and co-develop a mobile AR serious game (*LINA*) to address real-world social connection and sense of belonging in the classroom with and for early adolescents (aged 10+). *LINA* is a shared, immersive, interactive narrative in which players collaborate as they use augmented reality to uncover artefacts left behind by a fictional classmate, in order to determine, collectively, where she has gone and why. The game pursues two outcomes. First, it aims to strengthen sense of belonging by bringing together classmates to share both immersive and collaborative gameplay in a structured way, creating positive experiences around successfully solving puzzles in pairs or groups. Second, it aims to create a safe space for a class and their teacher to learn about and reflect on prejudice, stigma, and social wellbeing, especially around mental health. This leads to reduced stigmatisation and marginalisation of peers who present or behave differently.

In this paper, we describe the concept of *LINA*, including theoretical background, storyline and gameplay, and design and (co-)development processes, and present results from an initial evaluation study, where we quantitatively evaluated usability and qualitatively evaluated efficacy of the game.

## 2 RELATED WORK

Gamified interventions and serious games are increasingly used to train and teach users in various settings such as healthcare [25] or education [26] and have also been implemented successfully in a therapeutic setting [27], for example as a treatment for depression [28]. Positive outcomes have also been found when looking at serious games for children and adolescents: in their systematic reviews, Zayeni, Raynaud and Revet [29] show that serious games can be an effective tool in psychotherapy, while Balaintharanathan, Palmer, Arellano and Arcand [30] found that serious games are effective especially at imparting knowledge, but also at changing behaviours.

Few serious games have sought to improve social skills, especially in a classroom setting and for a general population. Existing social skills games have largely been developed for a target group of children with autism spectrum disorder (ASD) [for an overview, see 31], but most of these games are single-player, or played in pairs or with a facilitator. Some other games have been developed for a therapeutic setting. In SIDES [32], a cooperative tabletop computer game, students (in groups of four) learn group work skills such as perspective-taking or negotiation skills in a social group therapy setting.

AR learning applications are primarily used in the field of natural science [33, 34]. Again, AR games that address social skills are mostly targeted at children with ASD. In a systematic review about AR learning applications for children and adolescents with ASD from Khowaja, Banire, Al-Thani, Sqalli, Aqle, Shah and Salim [35], 20 out of the 30 analysed studies had outcomes related to social skills and competences.

The unique characteristics of AR to implement digital content in the real physical environment of the player allow for an exciting new direction in digital learning – a blended approach, where players engage both digitally and through 'real life' interaction with each other and the AR components. This is especially suited to addressing sense of belonging via a digital setting while still implementing real-world contact. One example of how this was done in a commercial game is *Pokémon Go*, where players come together and form groups in the real world, in order to fight in digital arenas. However, *Pokémon Go* players are not required to communicate or have real-world social interactions in order to achieve their game goals. It has been shown that AR can encourage cooperation and collaboration [23] and that collaborative tasks with AR may support learning outcomes [36]. However, in existing serious games, collaboration is mostly used as a method to reinforce specific learning outcomes, rather than being an outcome itself. For example, in the serious game *Pathomon* [37], players use AR with the primary aim of learning about viruses, but also collaborate by sharing with each other the correct ingredient combination for an antidote. However, players did not feel that their actions influenced the actions of others, which demonstrates an opportunity for greater emphasis on collaboration.

The importance of sense of belonging in early adolescence and the lack of interventions aimed at improving belonging led to the development of the presented game *LINA*. While collaboration is often used in AR games to achieve other learning goals, few games have tried to improve sense of belonging as a main outcome, especially targeting an entire group of people at the same time. Thus, our smartphone social AR game uses a blended approach of real-world and digital interactions to improve sense of belonging in early adolescents.

## 3 DESIGN AND DEVELOPMENT OF THE GAME *LINA*

### 3.1 Concept

*LINA* is a mystery-style interactive narrative experienced by an entire class (and their teacher) together. Players learn that a fictional classmate called Lina has suddenly left the school. Gameplay involves solving the mystery of where Lina – whose behaviour often marked her out as unusual – has gone, and why. Players uncover personal artefacts Lina has left behind in the classroom (using AR), exchange information with other players, and collaboratively solve puzzles. First alone, then in groups and finally as an entire class, they piece together the story of a classmate dealing with an atypical and stressful home life, caring for a mother suffering from depression.

The game has been developed in both English and German; co-development was with early adolescents in Austria. The name *LINA* was chosen for its acronymic potential in both languages – 'Lina Is Not Alone' / 'Lina Ist Nicht Allein'.

*3.1.1 Theoretical background of the game LINA*. Increasing sense of belonging and reducing barriers between individuals and groups is a concept primarily addressed in Contact Theory and stigma research [38-45]. Contact Theory argues that interaction involving (1) equal status, (2) cooperation, (3) a common goal, and (4) authority support, enhances social interactions and reduces prejudice between people, especially those perceived as outsiders or belonging to outlier groups [46]. Based on this, *LINA* has been designed so that (1) all players (including the teacher) start with the same amount of information; (2) information needs to be shared via gamified cooperative tasks and real-world interactions; (3) players are ultimately steered towards solving the mystery around Lina together; and (4) the game leads players safely and confidently through the game, offering appropriate support (e.g. in form of hints). Imagined and vicarious contact have also been shown to have effects [46]; thus, imagined contact with an individual affected by mental illness can lead to improvements around stigma, attitudes and social distance.

Exposure to the arts in general [47] as well as principles from drama and movement therapy [48, 49] are potent tools to improve mental health. Storytelling in particular has been used successfully to reduce prejudice and stigma, improving attitudes towards characters with 'outsider' status [50, 51]. *LINA* therefore immerses players in a story about mental ill health, not only educating them, but letting them play an active role in the story.

In summary, *LINA* has been designed for school classes to spend time together and improve their sense of belonging in the whole group. While scripted peer interactions (role-play) and immersive storytelling provide a space for safe communication and reflection on mental health, gamified collaborative tasks promote shared positive emotions, breaking down barriers between peers.

*3.1.2 Creative co-development with young people*. The game was co-developed with our target group of early adolescents throughout the development process. All workshops with schools were approved by the Karl Landsteiner University Commission for Scientific Integrity and Ethics (EK NR. 1025/2020).

At the start of development, we held a series of four exploratory workshops with two classes (grade 1, age 10-11) in Austria over a period of two months. The aims of these workshops were to (1) gain factual information about day-to-day life in grade 1 of a secondary school, as this was where *LINA* would be set; (2) gain information about the cultural lives of early adolescents such as likes and dislikes, activities, media, attitudes and language; (3) gain information about social dynamics and friendships in their classes, as well as causes and enactments of social conflict; and (4) explore the impact and consequences of unusual or transgressive behaviour within these groups. Further, students were introduced to the concept of the game to assess its attractiveness to them and seek contributions towards storyline and game execution. Information during those workshops was gathered via group or whole-class discussions, but also by having students draw, write or act out ideas for the game. Lastly, we invited adolescents to write private, anonymous letters both to the game development team and directly to Lina, to give them the opportunity to share thoughts and ideas they might not wish to share publicly. Insights from these workshops strongly influenced development of the game. The resulting collection of artwork, for example, helped the lead artist set a style for the game (see Fig. 1 and Fig. 2 for examples of how student drawings from co-development were translated into the game).

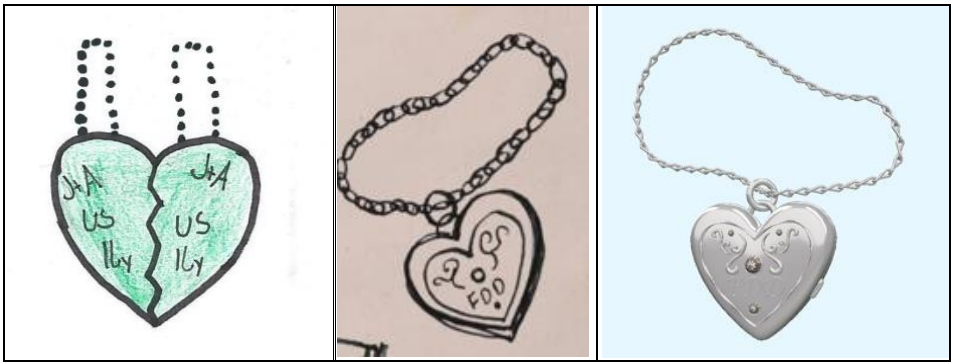

Fig. 1. Locket – Student artwork (left), artwork for the game (middle), AR object in the game (right).

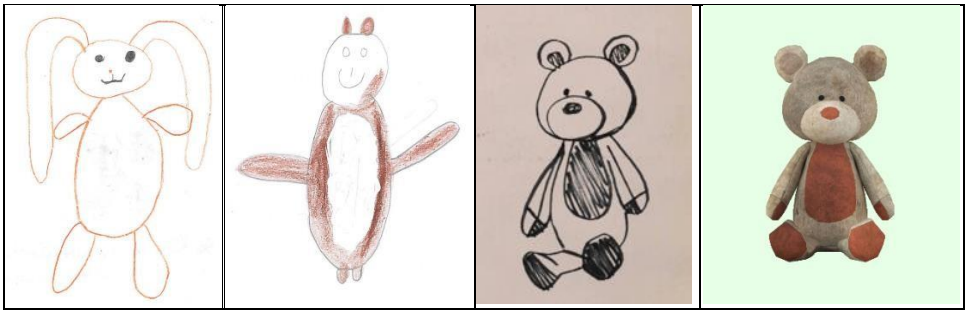

Fig. 2. Teddy bear – Student artwork (left and middle left), artwork for the game (middle right), AR object in the game (right).

In addition to co-development with adolescents, we conducted focus groups with four female adult participants who grew up with a parent with mental illness, to ensure our story felt truthful and reflected their reality. The aim of these conversations was to find out about the impact of having a parent with mental illness on their lives as early adolescents, exploring personal stories of incidents at school and the school-home dynamic. The main insights gained from these focus groups were that there was not one generic lived experience but many and varied specific experiences, which were often quite distinct; that young people in this situation do not necessarily recognise their home life as different because it's the only one they know; and that they instinctively seek to protect and normalise their parent(s) and parenting. These insights, as well as ideas for incidents and situations, influenced the development of the storyline around Lina and her mother.

*3.1.3 Storyline – De-stigmatising mental health problems.* Lina, who joined the school midway through the academic year, had not moved past newcomer / 'outsider' status when, just as suddenly, she departed. The class is left with an enigma around Lina – who she really was, where she's gone, and what to make of her short time with them. But Lina has pre-empted this: each player finds a (virtual) notepad underneath their desk that Lina has left for them, which reveals a series of puzzles and challenges she has set around how her classmates saw her. In absence, Lina leads the players on a journey of discovery around their classroom, finding (virtual) objects and messages that she has deliberately left behind. Each artefact prompts fresh memories around how students (including the player's fictional character) experienced and interacted with Lina, what kind of person she was, and what happened during her brief time in the class – for example, Lina's conflict with the teacher when she came to school wearing inappropriate shoes, or her guardedness around her diary, which could only be opened with a key kept on a string around her neck. The teacher, too, discovers scraps of information about Lina, though these raise more questions than answers – questions their students will be able to answer.

By the end of the game, players have discovered that Lina lived with just her mother, and that the two of them frequently moved, in part as a consequence of her mother's mental health issues. Some of Lina's unusual behaviours in class can be attributed to this life circumstance. For example, she was cautious about making friends because she anticipated moving again; she was reluctant to allow people to see where she lived; she had responsibilities towards being prepared for school (e.g. having the right kit) that other children do not have. At one point she attended parents evening herself, in lieu of her mother. At the end, it emerges that Lina's mother has realised that their situation is neither sustainable nor fair on Lina, and that they have moved one final time, to live with Lina's aunt and cousins while her mother seeks long-term treatment.

*3.1.4 Gameplay – endorsing sense of belonging.* Gameplay, which lasts up to one hour, begins with every student sat in their usual place. Each player has a smartphone with headphones attached, through which they hear an introduction about themselves: they are a fictional student, in a classroom just like the one they are playing in. The teacher (who has also been introduced to their fictional alter-ego) begins the morning register in a guided roleplay scene, where smartphone screens tell players what to say and when (Fig. 3). It becomes apparent that Lina (who is not played by any of the players) has been removed from the register. The teacher does not know where she has gone or why.

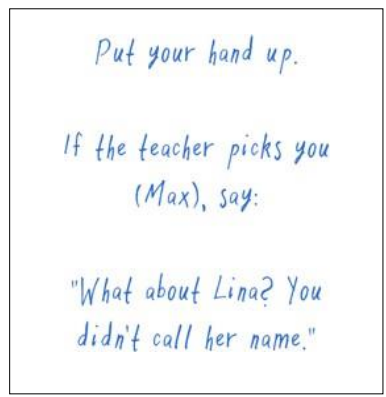

Fig. 3. Roleplay prompt in the introductory sequence of LINA.

Players then receive a message from Lina, telling them to look under their desk, where they discover, through the smartphone camera, a personalised (digital) notepad (using floor-detecting technology and AR) which guides them through the game (Fig. 4). The notepad directs players towards specific physical (printed) markers which have been preset around the classroom. Scanning the correct marker reveals an artefact belonging to Lina (AR object, see Fig. 5) which prompts a voiceover in which the player-character recalls incidents involving Lina's time in their class related to that object.

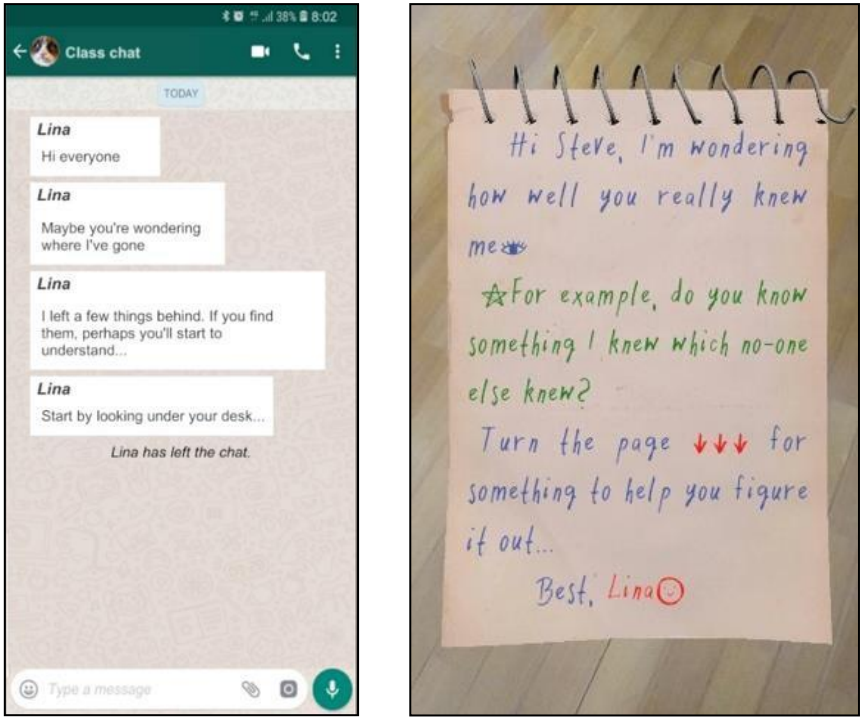

Fig. 4. Screens with messages from Lina: WhatsApp message (left) and notepad (right).

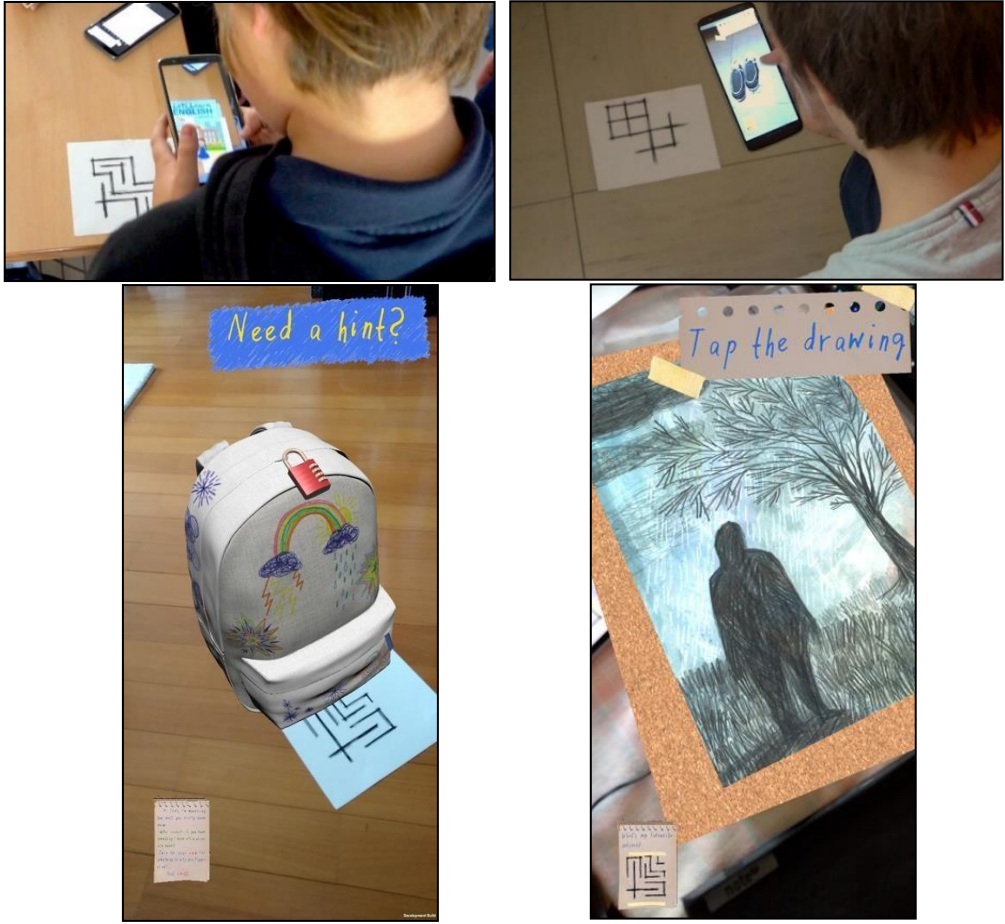

Fig. 5. Finding objects with AR in *LINA*.

Players learn different information about Lina depending on which markers they have been directed to scan. In the next part of the game, players are directed to find a specific partner, who will have acquired different information. Contact happens both on a digital and a physical level. The two players prove they have found each other by touching their phones together (using Near-Field Communication, NFC, Fig. 6). They then cooperate to solve a puzzle involving sharing information about the objects they have found (Fig. 7), so that by working together, each player finds out more about Lina.

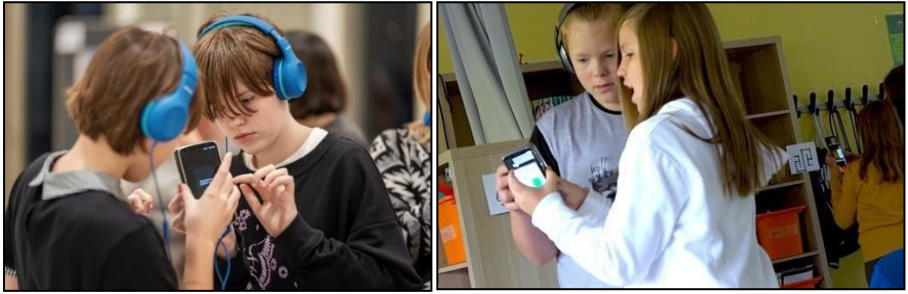

Fig. 6. Two players touching their phones to connect in-game via NFC.

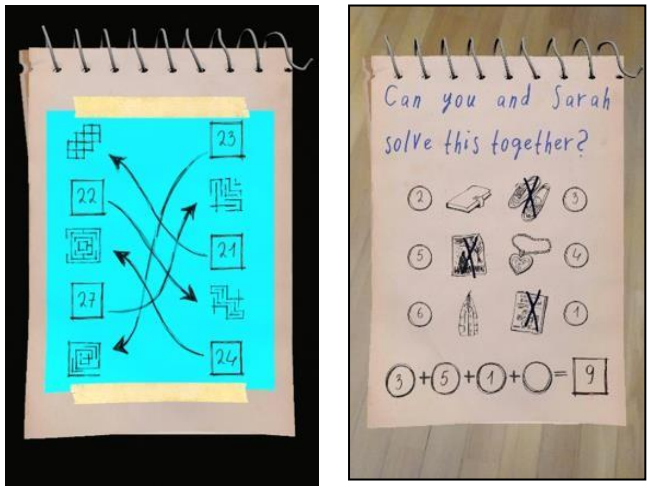

Fig. 7. A pair of players who need to collaborate to solve a puzzle. They work on Player B's screen (right) to determine which objects they have found. The resulting number unlocks the next marker to find on Player A's screen (left).

After pair interactions, players are directed towards new groups of three to four people (again, using NFC) to solve another puzzle; this leads the group to the teacher, with whom they share what they've learnt and who shares a further piece of information with them. Once each group has talked to the teacher, they take part in a fun challenge, tasked with finding everymarker hidden in the classroom within a set time. Player then discover a piece of paper (floor- detecting technology and AR) on the floor – a page of Lina's diary. Directed to form a circle,each player reads out their fragment of diary (Fig. 8), bringing together the remaining information and so solving the mystery around why Lina has left. After this, everyone is askedto return to their desks and be themselves again. In a guided post-game discussion, the teacher's smartphone offers a selection of discussion questions for the class to reflect upon (Fig. 9).

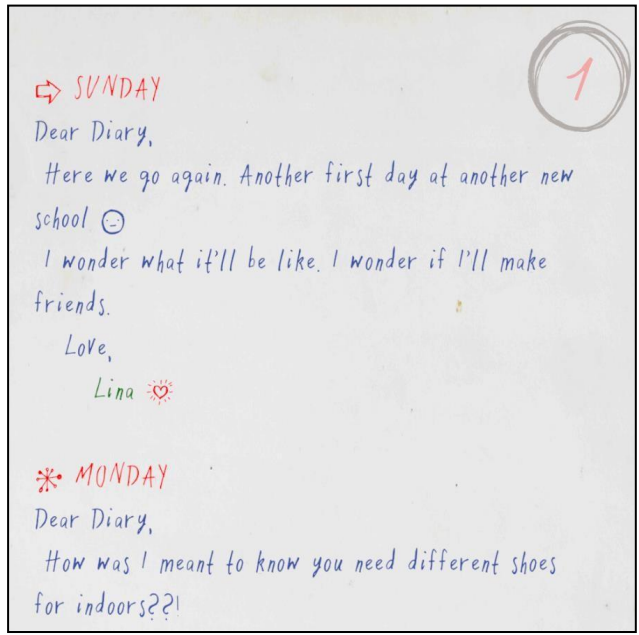

Fig. 8. One part of Lina's diary to be read out by a player.

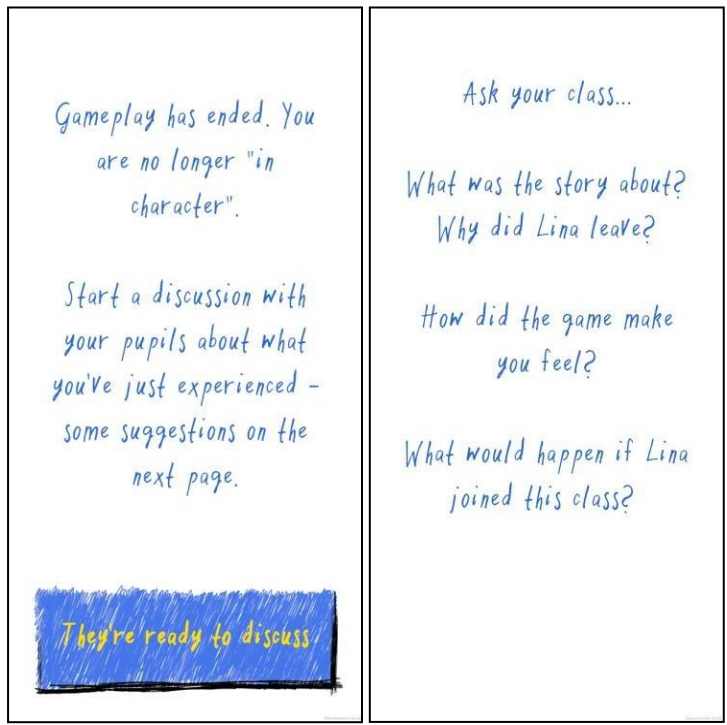

Fig. 9. Teacher's screen with post-game discussion prompts.

**3.2 Development**

*3.2.1 Interdisciplinary and international collaboration.* The pan-European interdisciplinary research group D.O.T. was set up to address social connectedness in early adolescents as a protective factor in mental health, exploring how arts-informed digital interventions could support social wellbeing. *LINA* is the product of an interdisciplinary and international collaborative process between a variety of disciplines and countries, with a focus on the interplay between technical development, arts, and psychology. The game is written and directed by a professional playwright, theatre director and creative practitioner from the UK, who worked closely together with Austrian psychologists, and computer scientists from Portugal and the UK, to implement storyline and psychoeducational content into the digital game. Additional AR programming and additional cooperation with industry partners such as a professional illustrator to create visuals (e.g. Lina's artwork and handwriting), and a professional composer for the soundtrack, ensured a polished audio-visual experience. Voiceover was performed by young people equivalent in age to the characters they were playing, plus adult actors to represent the teacher (see acknowledgements for further information).

Influences for *LINA* include immersive theatre (for example the work of the British theatre company Punchdrunk), participatory theatre and previous experiments around digital-theatre crossovers; walking simulator narrative games (e.g. *Gone Home*, *Firewatch*); and adolescent pop culture (e.g. the book and television series *13 Reasons Why*), encompassing the topic of mental health in the mystery narrative underpinning the game.

*3.2.2 Iterative co-development of prototypes.* After development began, different prototypes were tested with various classes at various time points. In an iterative process, ideas and feedback were integrated in each subsequent version of the game, thus continuously adapting the game to early adolescents' needs and ensuring that the world and story of the game felt convincing and compelling. In these workshops, we explored initial interactions with the smartphone, difficulty level (at a technical level, for game flow and understanding of storyline) and general opinions about the game with early adolescents (see acknowledgements for participating schools). Findings from these workshops influenced further development (e.g. refinement to puzzle design, refinement to UI and UX, implementing a time-delayed hint system for phases of gameplay which some players found difficult).

*3.2.3 Technical development.* *LINA* was developed in Unity[1] Engine, which provided the quick prototyping learning curve needed to create an initial proof-of-concept. Augmented Reality is provided by Vuforia[2] Engine in two different ways: Target Recognition uses an algorithm to identify images (markers) in the camera feed and place 3D models in a position related to the recognized marker; Ground Plane recognizes where the physical floor is and places digital objects there.

Given the requirement for multiplayer interaction, *LINA* follows a server-client architecture with two main components acting simultaneously: (1) The "LINA Master" application, running on a computer with Microsoft Windows, responsible for creating the game room, setting up, starting and managing the game, storing the state of the game and synchronizing the game for all client players; (2) The "LINA Client" app running on smartphones with Android operating system, where the players interact with the game itself. The communication between these two components is handled by the Photon[3] Multiplayer Game Engine for Unity, a service that provides a server hosting the game room.

---

[1] https://unity.com/, last accessed 21.02.2022

[2] https://developer.vuforia.com/, last accessed 21.02.2022

Beyond AR, the game uses Near-Field Communication (NFC), which allows two mobile devices to recognise and communicate with each other when in close proximity. One device registers as a sender, the other as a receiver; when devices are held back-to-back, a one-way message is sent. This is particularly useful in making sure two players have found each other. The game also takes advantage of the impact of multiple phones in a single contained space by, for example, using each player's phone as a separate audio speaker to create a polyphonic soundscape with different instruments placed in different parts of the room, or by creating synchronised 'choreography' of players by sending all players an instruction to move (e.g. stand up, or form a circle) at the same time.

Iterative development of *LINA* consisted of three major phases: (1) An initial proof-of-concept, which explored the augmented reality feature and modes of communicating with multiple players. (2) A first prototype, that incorporated placeholder 3D models and explored collaborative gameplay between the players. This prototype had no networking component and was completely offline. (3) The final prototype, with a complex multiplayer networking component and the introduction of new features like Near Field Communication, a checkpoint system to allow resumption of progress in the event of a technical problem, fully-animated 3D models, new music and narration sound files, and new multiplayer challenges.

## 4 EVALUATION STUDY

The aim of the evaluation study was to (1) quantitatively assess usability of the prototype of *LINA* with a questionnaire and (2) qualitatively assess usability and efficacy of the game by conducting focus groups. *LINA* was tested for both usability and efficacy during a series of workshops with early adolescents in June and July 2021. Five classes participated in the study, with a total of 91 participants between 10 and 13 years (age$_{mean}$=11.41; 47% female, 50% male, 3% missing data). The participants first played the game (~1 hour) and then filled out a questionnaire about usability of the game. Lastly, they discussed their experiences in small focus groups (~10-15 mins). The workshops were delivered by five trained workshop leaders, who set up the game, helped players during gameplay (e.g. with technical difficulties) and led the focus groups.

### 4.1 Materials and analysis

The German version of the Game User Experience Satisfaction Scale [GUESS-18; 52] was used to assess usability, feasibility and acceptability of the game. The questionnaire consists of nine constructs (two items per construct; 18 items in total): usability/playability, narratives, play engrossment, enjoyment, creative freedom, audio aesthetics, personal gratification, social connectivity, and visual aesthetics. The answers on the 5-point Likert scale range from 1 ("don't agree at all") to 5 ("totally agree"). Analysis was done with SPSS [53]. We analysed descriptive statistics with mean scores for all subscales. Considering there are still sex differences in the use of video games (only 36% of gamers were female in the fourth quarter of 2020, see [54]), we also considered sex differences in the analysis of the questionnaire. Due to missing normal distribution (tested with Kolmogorov-Smirnov and Shapiro-Wilk test), we analysed sex differences with the Mann-Whitney-U-Test.

---
[3] https://www.photonengine.com/pun, last accessed 21.02.2022

After playing the game, we conducted semi-structured discussions with small focus groups (3-4 per group). We aimed to qualitatively assess three main questions: (1) how participants enjoyed the general experience of the game, including elements like storyline, audio-visuals, difficulty level, and technical adaptation; (2) whether participants understood the storyline and what the game is trying to convey (de-stigmatisation); and (3) how participants experienced the cooperative parts of the game and whether they felt it helped their connectedness with their peers (sense of belonging). Discussions were recorded and later transcribed. We used principles of thematic analysis to group themes according to the research questions.

**4.2   Results and discussion**

Results of the GUESS-18 can be found in Table 1 (see Appendix for analysis on item level). Internal consistency was good with Cronbach alpha α= .83. The overall usability score was 33.64. The highest ranked construct was enjoyment (mean score=4.42), closely followed by social connectivity (mean score=4.37). The high scores in narratives, audio, and visual aesthetics might reflect the advantages of collaborating with industry professionals when creating a serious game. Looking at the medians of the subscales (see boxplots in Fig. 10), four subscales had a median value of 3.5. The lower scores in the subscale creative freedom could be expected, as the game is delivered in a structured way (including scripted roleplay and clear instructions) and was not intended to leave a lot of freedom during gameplay, though this is a decision that could be reassessed in future work. Engrossment scores might have been lower due to the fact that players are not supposed to be completely submerged in the game. One item reads, "I forget what is happening in the outside world while I play", but due to the blended approach, players are also playing in the 'outside world'. Lower gratification scores may also be due to the game being collaborative rather than competitive, though may also suggest more could be done to reward progress. Usability scores might have been affected by technical problems during gameplay of what was still a prototype of a complex piece of new software (see results of focus group discussions for further information).

There were no significant sex differences for the overall score, yet some subscales revealed sex differences. Female participants showed significantly higher scores in visual aesthetics, enjoyment and narratives, while males showed higher scores in usability/playability, which suggests that boys found the game easier to control while girls appreciated the overall experience, visuals and story of the game more. This shows that we succeeded in creating an experience that appeals to both sexes, but also indicates that males may have an advantage in controlling the game more easily, perhaps due to being more experienced in gaming, but may be less receptive to its narrative and aesthetics.

Although our aim was to assess usability (not efficacy) with the GUESS-18, the results also look promising in terms of expected efficacy and outcomes of *LINA*. The two highest-ranked constructs were enjoyment and social connectivity, which reflects our aim to bring students together and increase sense of belonging (reflected by the high score in the social connectivity subscale) in a fun way (reflected by the high score in the enjoyment subscale).

Table 1. Results from the GUESS-18

| Subscale | Mean M | Standard deviation SD | Mann-Whitney-U-Test (Sex differences) |
|---|---|---|---|
| Overall score | 33.64[a] | 4.87 | n.s. |
| Enjoyment | 4,42 | .84 | $U(N_f=43, N_m=45)=, z=-2.915, p=.004*$ (females show higher scores) |
| Social connectivity | 4,37 | .66 | n.s. |
| Narratives | 4,32 | .75 | $U(N_f=43, N_m=45)=, z=-2.057, p=.04*$ (females show higher scores) |
| Audio aesthetics | 4,07 | .96 | n.s. |
| Visual aesthetics | 4,02 | .71 | $U(N_f=43, N_m=45)=, z=-2.511, p=.012*$ (females show higher scores) |
| Usability/playability | 3,70 | .85 | $U(N_f=43, N_m=45)=, z=-2.323, p=.02*$ (males show higher scores) |
| Personal gratification | 3,46 | 1.01 | n.s. |
| Play engrossment | 3,42 | .98 | n.s. |
| Creative freedom | 3,32 | .97 | n.s. |

[a]Overall scores are calculated by adding up all subscales. Scores range from 9 to 45, see [52]

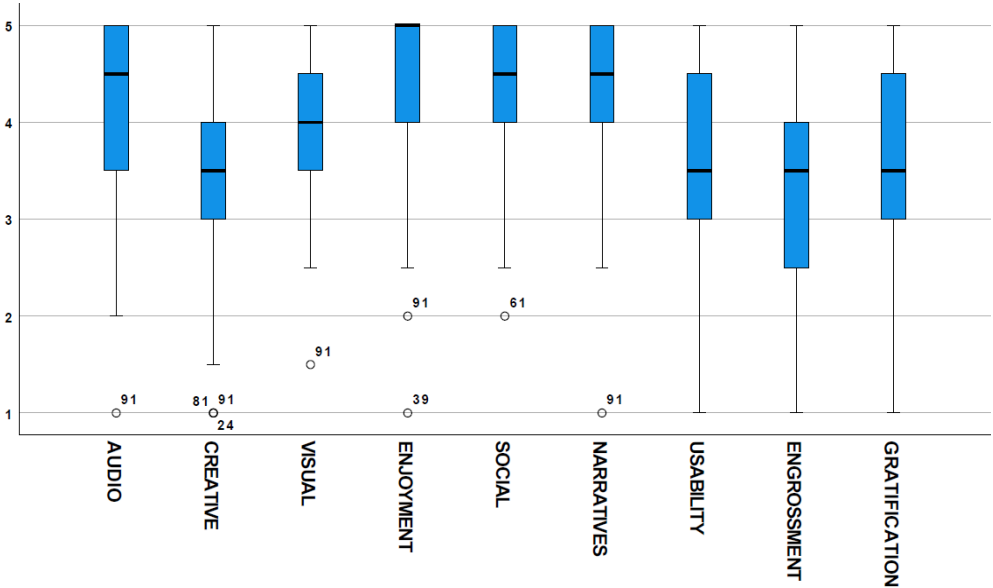

Fig. 10. Boxplot for the subscale scores of the GUESS-18: each plot shows one subscale (2 items).

For the focus groups, the feedback about the game was very positive in general. Almost all adolescents enjoyed playing it and would play this or a further episode of the game again. There was some critical feedback for specific parts of the game. The main problems adolescents had during gameplay were due to technical problems and waiting times (especially in the group phase, where some groups were faster with puzzle solving than others), which interrupted the flow and led to some players feeling bored. In terms of storyline, some wished for more explanation at the end, or mentioned what could be further explained in putative subsequent "episodes" of the game. Lina's home life, her relationship with her mother, and how her life is now that she has moved back were topics of interest. However, adolescents realised that the storyline was about de-stigmatising (see quotes in Table 2). Answers relating to sense of belonging showed that adolescents were aware that the main aim of the game is to bring players together and improve class climate, and also thought that the game was able to achieve that (see Table 2).

Table 2: Example quotes from adolescents from focus groups discussions

| Topic | Example questions asked by the researchers | Quotes from adolescents |
| --- | --- | --- |
| General | How did you like the game? | *R: How did you like it? Tell me!*<br>*A1: Nice, very nice.*<br>*A2: Very good, very very good.*<br>*A3: 100 out of 10!*<br>*A4: I say 8 out of 10.* |
| | What did you like/dislike? | *Sometimes it was complicated when something didn't work, but apart from that I really liked it.*<br><br>*The only difficult thing was that we had to wait for everyone, because they needed more time for something.*<br><br>*When I heard about it, I thought that only the children were playing and I thought it was cool that the teachers were also playing.*<br><br>*I also thought it was cool that we got the names, because then we also played the role, that we had to stand up in the classroom and say "Good morning, Mr. Gruber".* |
| Sense of belonging | What do you think was the purpose of playing *LINA* with your class? | *A: That we build up more contact with the class.*<br>*R: Did that work?*<br>*A: Yep* |
| | What, if anything, do you think *LINA* has helped you and your class with? | *I think this has also strengthened the class community a bit because we all worked together.*<br><br>*So, you play it together in the class. Because then everyone worked closer together and that was also* |

416| Topic | Example questions asked by the researchers | Quotes from adolescents |
|---|---|---|
| De-stigmatisation | What, if anything, do you think *LINA* has helped you and your class with? | *fun and that you simply do more together as a class.* <br><br> *It's also fun, because you also do something with those you actually do not have so much contact with* <br><br> *I think with Lina it was also a little bit that you learn that you do not have to exclude people and should not.* <br><br> *R: And why do you think we are here today to play the game with you?* <br> *A: So that the kids who play it realize that they shouldn't bully others. Or judge them for anything.* |
| | What was the story about? | *And Lina has left all these items that if you have solved the puzzle you can also understand why she had to leave now and in any case I really liked the game and it was very great.* |

## 5 CONCLUSION

In this paper we introduce *LINA*, a smartphone AR serious game that is played by a whole class of early adolescents to reduce stigma around mental health and improve sense of belonging. The game was developed by an interdisciplinary and international team and involved early adolescents as co-developers throughout development. Both quantitative and qualitative data from the evaluation study showed high acceptability of the game, and qualitative data show promising results regarding efficacy. A limitation of the study is that the data for efficacy are only preliminary due to their qualitative nature. We plan to conduct a further evaluation study to quantitatively test efficacy of the game.

A major strength of *LINA* lies in the interdisciplinary co-development process: collaborating with professional industry partners especially for narrative, visuals and audio increased the chance of a high-quality game. Involving early adolescents throughout the development process created a game that reflects the voices of its target group.

For future work, co-development left us with an abundance of material and ideas that could still be implemented, suggesting that gameplay and narrative could be sustained across multiple episodes. While the current version of the game is played in one session, there is the potential to rework and expand *LINA* as an intervention lasting several weeks, playing a new episode each week.

The impact of *LINA* may surpass the current game, in that it showcases the general potential of this form of social AR gameplay in educational, psychotherapeutic or entertainment settings – a collective experience that fosters a greater sense of belonging while creating a safe environemnt in which to engage players in a potentially challenging topic. The fact that an increasingly ubiquitous digital device such as a smartphone acts as the facilitator of the game makes it cost-effective: easy to distribute digitally, and easy to implement without externalfacilitators. The affinity of adolescents for digital media makes them an especially appropriate target for this kind of intervention; the format might also be used for building sense of belonging in other environments, for example in a workspace with a group of adults or withany other group that wants or needs an entertaining way to connect both digitally and in thereal world. In that sense, beyond being a stand-alone intervention, *LINA* could be seen as a prototype for "Augmented Social Play" interventions, which could be adapted to the specific needs of the target group while always endorsing sense of belonging.

## A  APPENDIX

Table: Results of the GUESS-18 on item level.

| Items (Audio, Creative, Visual, Enjoyment, Social, Narratives, Usability, Gratification, Engrossment) | Mean | Min | Max | Standard Deviation | Variance |
|---|---|---|---|---|---|
| I like the sound effects in the game. | 3,98 | 1 | 5 | 1,14 | 1,30 |
| I think the sound in the game (e.g. sound effects and music) improves my gaming experience. | 4,19 | 1 | 5 | ,95 | ,90 |
| I think the game allows me to make changes. | 3,59 | 1 | 5 | 1,09 | 1,18 |
| I think I can be creative while playing the game. | 3,06 | 1 | 5 | 1,14 | 1,29 |
| I like the game's graphics. | 3,99 | 1 | 5 | ,89 | ,79 |
| I think the game looks nice. | 4,04 | 1 | 5 | ,81 | ,66 |
| I think the game is fun. | 4,52 | 1 | 5 | ,86 | ,74 |
| I am bored while playing the game. (re-coded) | 4,37 | 1 | 5 | 1,05 | 1,10 |
| I think the game promotes communication (e.g. talking, chatting) between players. | 4,38 | 2 | 5 | ,71 | ,51 |
| I like playing the game with other players. | 4,36 | 1 | 5 | ,90 | ,82 |
| The story of the game got me excited from the start. | 4,25 | 1 | 5 | ,82 | ,68 |
| I like the story of the game. | 4,43 | 2 | 5 | ,77 | ,59 |
| I find the game easy to control. | 3,66 | 1 | 5 | ,98 | ,96 |
| I find the in-game interface easy to use. | 3,73 | 1 | 5 | ,94 | ,90 |
| I want to do as well as possible during the game | 3,50 | 1 | 5 | 1,20 | 1,44 |
| I am very focused on my own performance while playing the game. | 3,46 | 1 | 5 | 1,11 | 1,24 |
| Sometimes I lose track of time while playing the game. | 3,69 | 1 | 5 | 1,06 | 1,12 |
| I forget what is happening in the outside world while I play. | 3,16 | 1 | 5 | 1,27 | 1,61 |

21neurohormones in adolescents with mild depression. International journal of neuroscience, 115, 12 (2005), 1711-1720. https://doi.org/10.1080/00207450590958574

[50] Lindsey Cameron and Adam Rutland 2006. *Extended Contact through Story Reading in School: Reducing Children's Prejudice toward the Disabled*.

[51] Lindsey Cameron, Adam Rutland, Rebecca Douch and Rupert Brown. 2006. Changing Children's Intergroup Attitudes toward Refugees: Testing Different Models of Extended Contact. Child Development, 77, 5 (Sep-Oct 2006), 1208-1219. https://doi.org/10.1111/j.1467-8624.2006.00929.x

[52] Joseph R Keebler, William J Shelstad, Dustin C Smith, Barbara S Chaparro and Mikki H Phan. 2020. Validation of the GUESS-18: A Short Version of the Game User Experience Satisfaction Scale (GUESS). Journal of Usability Studies, 16, 1 (2020),

[53] IBM Corp. 2016. *IBM SPSS Statistics for Windows, Version 24.0*. IBM Corp, Armonk, NY.

[54] Statista *Gender distribution of gamers worldwide Q4 2020, by device*. J Clement, City, 2021.